# Graph Model Implementation of Attribute-Based Access Control Policies

Hadi Ahmadi and Derek Small
Nulli Identity Management, Calgary, Canada
{hahmadi,dsmall}@nulli.com

**Abstract.** Attribute-based access control (ABAC) promises a powerful way of formalizing access policies in support of a wide range of access management scenarios. Efficient implementation of ABAC in its general form is still a challenge, especially when addressing the complexity of privacy regulations and access management required to support the explosive growth of social and IoT networks. In this paper, we introduce a graph model implementation for expressing and evaluating access policies and illustrate a sample use-case implementation over Neo4j™ Graph Database. Graph databases excel at querying connected data and hence can evaluate complex policies efficiently via graph traversal algorithms.

## 1. Introduction

In attribute-based access control (ABAC) [1], the level of access to resources are determined using attributes defined over the access request including user, resource, environment parameters, etc. This model allows for enforcing complex access control policies that involve arbitrary combination of attributes with static, dynamic, and relationship-based values, thus extending beyond role-based access control (RBAC) [2] and relationship-based access control (ReBAC) [3].

While ABAC allows for high granularity of access policies and hence fits a wide range of access management requirements, implementing ABAC is generally a challenge as it may need more time and effort for setup, deployment, and runtime processing. More importantly, the average processing time increases with growing policy flexibility and complexity. A variety of implementation models have been studied over time; however, they either have been specific to a use case, e.g. web services [4,5], or have considered hybrid approaches [6,7] that sacrifice the flexibility of “pure” ABAC. One may argue that eXtensible Access Control Markup Language (XACML) [8] is the widely-accepted standard model for ABAC; we note, however, that XACML is an access control policy language that has support for attributes, hence to be viewed as a potential component of an ABAC model.

This paper explores formalizing an ABAC graph model that promises performance of runtime processing while allowing for ease of policy administration and maintenance over time. The key to our graph model implementation is allowing the ABAC primitives, attributes, and policies to coexist as part of a single graph structure. In precise words, primitive data parts such as subjects, objects, actions as well as their attributes are presented as graph nodes that are connected through conceptual relationships. ABAC policies are defined as authorization subgraphs that connect to attributes/primitives through condition relationships. This allows access management systems to make policy decisions based on well-established and efficient graph traversal algorithms. Thanks to the evolution of graph databases, we can implement our ABAC graph model without being worried about dealing with relational databases and their inherent JOIN operations. We particularly use Neo4j Graph Database [9] for the purpose of illustrating our use-case examples.

## 1.1. Related Work

A variety of implementation models for ABAC have been studied recently, most of which aim at generalizing traditional access control models. Wang et al. introduced a logic-based framework for ABAC [10]. Zhang et al. proposed use of attribute enhanced access matrices [11]. More recently, Rubio-Medrano et al. [12] put forth use of security tokens and Ferraiolo et al. formalized the notion of “policy machine” for ABAC [13]. Jahid et al. [14] developed MyABDAC relying on relational database capacities and XACML policy language. The idea of using graph implementation for ABAC has been investigated by Santiago et al. [15] and, more recently, by Jin and Kaja [16]. Despite similarities, the implementation approaches in [15,16] do not consider a unified model to design policy graphs and hence cannot offer universal graph traversal algorithms for policy decisions.

## 1.2. Notations

We base our notations on Cypher-query language: “Cypher is a declarative graph query language that allows for expressive and efficient querying and updating of the graph store” [17].

We use $(:X)$ (resp. $[:X]$) to denote the set of all nodes (resp. relationships) of type $X$. The set $(:X)\times(:Y)$ denotes the cartesian product of two sets; similarly, $(:X)^n$ means $n$ times cartesian product of $(:X)$ with itself. The union of sets $(:X)$ and $(:Y)$ is shown by $(:X|Y)$. Similarly, $[:X|Y]$ denotes the union of relationship sets $[:X]$ and $[:Y]$. The notation $[:Y*m..n]$ shows all paths of length $m$ to $n$ with relationships of type $Y$.

## 1.3. Organization

This paper is organized as follows. In section 2, we describe a condition-based representation of ABAC which helps us introduce our core concepts and our graph model in Section 3. Section 4 illustrates an example access scenario and examines our graph model implementation. We conclude the work and suggest directions for future work in Section 5.

# 2. Condition-Based Representation of ABAC

We consider a representation of Attribute-Based Access Control (ABAC) where access control policies are defined as “Permit” or “Deny” decisions over a set of Condition predicates (or simply Conditions). Access control policies are created and managed to restrict access to protected objects. An access request contains the following main primitive primitives (or “primitives” for short):

- subject
- action
- object

In a simple explanation, an access policy can be presented as (***Permit/Deny: Conditions***), implying:

***Permit/Deny*** *subject to perform an action on an object if* ***Conditions*** *hold.*

Each condition involves subjects, objects, actions, environment parameters and relationships between them. Examples of condition clauses are:

- *subject* belongs to the group “Admin”.
- *object* is of file type “PDF”.
- *action* is part of the process “Manage User”.
- *subject* has “Owner” *relationship* with the *object*.

**Note:** In this paper, we focus on “simple” “monotonic” policies where each policy condition is instantiated by either a primitive (e.g., “jsmith”) or an attribute (e.g., “Admin”). In more complex ABAC scenarios, policy conditions can be compound and include “disjunction”, “conjunction”, and “negation” of other conditions. We also stay with “subject”, “object”, and “action” as our query primitives and do not consider other attributes including environmental parameters. In “Conclusions and Future work” section, we discuss some challenges of implementing compound ABAC and provide examples of how our solution can be expanded.

# 3. Graph Model Implementation of ABAC

We describe our model by defining the main concepts and components, followed by introducing our universal decision making and policy combining algorithms.

## 3.1. Primitive and attribute nodes

Modeling organizational data into graph can proceed in various ways depending on the use cases defined around the data. From the ABAC point of view however, it suffices to distinguish between “primitive” elements, i.e., subjects, objects, and actions, and the types of “attributes” defined for them. Hence, in our graph model of ABAC we categorize organizational data nodes as “primitive” nodes and “attribute” nodes:

- ***A "primitive" node*** is a node in the graph that refers to an individual subject, action, or object.
- ***An "attribute" node*** is a node that describes a primitive node via a direct relationship or a chain of attribute-node relationships.

In our graph model, we use $[: HAS_ATTR]$ *relationship to connect* primitive nodes with attribute nodes and attributes nodes with one another. For example, the Cypher pattern:

$$(x : Primitive) - [: HAS_ATTR] -> (att1 : Attribute) - [: HAS_ATTR] -> (att2 : Attribute)$$

shows a chain of attributes for the primitive node $x$.

## Policy and condition nodes

Each access control policy over graph is realized as a set of paths that represent condition clauses. Put differently, we represent a "policy subgraph" with a uniquely identified policy node "$pol$", with label $:Policy$, properties including $decision \in \{Permit, Deny\}$, and a set of condition relationships $\{con_i\}_{i=1}^{*}$ that connect "$pol$" to condition nodes $\{c_i\}_{i=1}^{*}$ in the graph. We define the array of possible condition types as

$$CT = (SUB_CON, ACT_CON, OBJ_CON),$$

where

- $SUB_CON$ ⇒ representing a subject condition
- $ACT_CON$ ⇒ representing an action condition
- $OBJ_CON$ ⇒ representing an object condition

This implies that every policy node is connected to the rest of the graph via relationships of type $con_i \in CT$. For a given policy node "$pol$", its policy subgraph can be obtained by the Cypher pattern:

$$(pol) <- [: SUB_CON|ACT_CON|OBJ_CON] - (c)$$

A condition node "$c \in (: Primitive|Attribute)$" can be a primitive node or an attribute node that connects to the primitives or their relationships. This allows us to define access control policies for specific subjects, actions, objects, as well as their relationships and attributes.

We assume that every policy node has at least one "$SUB_CON$", one "$ACT_CON$", and one "$OBJ_CON$" relationship; otherwise, it is invalid.

**Definition 1 (Valid Policy).** *A policy node "$pol$" is called <u>valid</u> if for all $con \in CT$, the Cypher pattern:*

$$(pol) <- [: con| - (c)$$

*is non-empty.*

**Note:** *For the rest of this paper, we assume all policies are "valid" unless otherwise mentioned.*

Furthermore, we treat each policy node as an AND operator for all the connected conditions, i.e., all conditions should meet for a policy to match. This is explained formally in Definition 4.

## 3.2. Access queries and policy matching

The main strength of our ABAC graph model is in allowing the Graph to be self contained for authorization queries. That is data and policies sit together in the graph DB and hence one universal graph traversal algorithm can be used to find access control decisions.

Generally speaking, the decision query looks for patterns that connect policy nodes to the system-given subject, object, and action nodes as “access query” primitives.

**Definition 1 (Access Query).** *An Access Query is a triple* $AQ = (sub,\ act,\ obj) \in (:Primitive)^3$*, where*

- $sub$ *identifies the subject who requests access*
- $act$ *identifies the type of access by the subject over the object*
- $obj$ *identifies the object (resource) that is requested to be accessed*

*Each of* $\{sub,\ act,\ obj\}$ *is called a query primitive.*

To determine whether a policy is relevant to an access query, we need to obtain what that policy requires as conditions.

**Definition 2 (Required Condition).** *An attribute/primitive node* $c \in (:Primitive|Attribute)$ *is called a* <u>*required condition of type*</u> $con \in CT$ *for the policy “*$pol \in (:Policy)$*”, if and only if:*

$$(c) - [:con] -> (pol)$$

*is non-empty.*

Next, it is important to obtain whether the required conditions are satisfied by the access query.

**Definition 3 (Satisfied Condition).** *A condition node “*$c$*” is* <u>*satisfied by*</u> *query primitive “*$x$*” if and only if:*

$$(x) - [:HAS_ATTR * 0..N] -> (c)$$

*is non-empty, where “N” is the “Graph Attribute Depth”, i.e, the maximum length of attribute chains.*

Now, we can define what a matching policy for a query should look like.

**Definition 4 (Matching Policy).** A policy node “$pol$” matches a given access query $AQ = (sub,\ act,\ obj)$*, if and only if each required condition “*$c$*” of type “*$CT_i \in CT$*” is satisfied by query primitive* $x = AQ_i \in AQ$*. For example, if “*$c$*” is of type “*$SUB_CON$*”, it should be satisfied by “*$sub$*”.*

## 3.3. Policy combining and decision making

The last step of policy evaluation is making decisions based on matching policies. It is obvious that if all matching policies are of "Permit" effect (i.e., include $decision = Permit$ property), then policy combining will be as simple as the existence a matching policy. However, this is not always the case as in certain scenarios, we need to deal with both Permit and Deny matching policies and it becomes crucial to define a procedure for arriving at a decision when there is conflict. Policy combining algorithms play an important role in such situations.

**Definition 6 (Policy Combining Algorithm).** *A policy combining algorithm* $\lambda$ *is defined as a boolean function* $\lambda : (:Primitive)^3 \times (:Policy)^* \rightarrow \{Permit, Deny\}$ *that for a given access query* $AQ$ *and a set of matching policies* $\{pol_i\}_{i=1}^*$ *returns an access decision.*

Common policy combining algorithms include "deny-overrides", "permit-overrides", "first-applicable", etc. More interesting examples of such algorithms, when it comes to graph traversal, can involve using graph characteristics such as shortest path, policy node properties, and policy path weight.

## 3.4. Universal decision statement

We have so far provided a graph model implementation of ABAC, which suggests how to create policy subgraphs and later how to obtain policy matching and decisions for runtime access queries. In this section, we conclude our model by presenting a universal Cypher statement that returns the matching policies and the ultimate decision for a query.

**Claim 1 (Policy Matching Query).** *The following Cypher statement (where* $N$ *is the graph policy depth) returns "all" valid matching policies for access query* $AQ = (sub, act, obj)$.

```
match (sub)-[:HAS_ATTR*0..N]->(sc)-[:SUB_CON]->(pol:Policy)
with pol, size(collect(distinct sc)) as sat_cons
match (pol)<-[:SUB_CON]-(rc)
with pol, sat_cons, size(collect(rc)) as req_cons where req_cons = sat_cons

match (obj)-[:HAS_ATTR*0..N]->(sc)-[:OBJ_CON]->(pol)
with pol, size(collect(distinct sc)) as sat_cons
match (pol)<-[:OBJ_CON]-(rc)
with pol, sat_cons, size(collect(rc)) as req_cons where req_cons = sat_cons

match (act)-[:HAS_ATTR*0..N]->(sc)-[:ACT_CON]->(pol)
with pol, size(collect(distinct sc)) as sat_cons
match (pol)<-[:ACT_CON]-(rc)
```

```
with pol, sat_cons, size(collect(rc)) as req_cons where req_cons = sat_cons

return pol as matching_policies
```

*Proof in Appendix C.*

**Note:** *The above query statement assumes that the “subject” node is the starting point to obtain policy matching over graph. The starting node however can change to “object” or “action” depending on which one leads to a better performance.*

## 3.5. Policy combining algorithms

Although Definition 6 suggests independent Cypher statements for policy matching and decision making, it is more efficient to obtain an optimized Cypher statement that incorporates our policy combining algorithm and returns the ultimate decision for any given access query. Unlike policy matching though, providing a general Cypher statement for decision making is challenging and depends highly on the type of policy combining. Below, we explain how we can convert the policy matching statement of Claim 1 to a decision making statement for a number of combining algorithms.

### *Deny/Permit-overrides*

The deny-overrides combining algorithm returns “Deny” when at least one of the matching policies has $decision = Deny$. We thus need to change the last line of the statement as (noting that lack of matching policy implies a “Deny”):

```
return case when count(pol) = 0 or 'Deny' in collect(pol.decision) then 'Deny' else
'Permit' end as decision
```

Similarly, the permit-overrides algorithm returns “Permit” when at least one of the matching policies has $decision = Permit$. The Cypher modification hence looks like:

```
return case when 'Permit' in collect(pol.decision) then 'Permit' else 'Deny' end as
decision
```

### *Max-score*

We define *max-score* algorithm as considering the policy node with the maximum $score$ (assuming policy nodes have an Number-valued property $score$). Since there can be multiple policies with the same $score$, we would need to combine this max-score with other metrics to complete our policy combining. For instance, a “***max-score-deny-overrides***” is obtained via the following return clause (assuming $pol$ set is not empty):

```
with max(pol.score) as max_score, pol where pol.score = max_score
return case when 'Deny' in collect(pol.decision) then 'Deny' else 'Permit' end as
```

```
decision
```

***Shortest-path***

The ABAC graph model implementation allows us to define graph based metrics. One example of such a metric is “shortest-path” policy, where a policy length is defined as follows:

**Definition 7 (Policy Length).** For access query $AQ$ *and policy* $pol$*, the policy length is defined as the sum of path lengths between* $pol$ *and each primitive* $x \in AQ$*.*

Although we can add the shortest-path logic to the end of policy matching statement, it will be more efficient to keep track of policy length while finding matching policies. Here is an example modification of the statement of Claim 1 to work as a “shortest-path-deny-overrides” algorithm (with modified sections show in bold):

```
match path=(sub)-[:HAS_ATTR*0..N]->(sc)-[:SUB_CON]->(pol:Policy)
with pol, length(path) as plen, size(collect(distinct sc)) as sat_cons
match (pol)<-[:SUB_CON]-(rc)
with pol, plen, sat_cons, size(collect(rc)) as req_cons where req_cons = sat_cons

match path=(obj)-[:HAS_ATTR*0..N]->(sc)-[:OBJ_CON]->(pol)
with pol, length(path) + plen as plen, size(collect(distinct sc)) as sat_cons
match (pol)<-[:OBJ_CON]-(rc)
with pol, plen, sat_cons, size(collect(rc)) as req_cons where req_cons = sat_cons

match path=(act)-[:HAS_ATTR*0..N]->(sc)-[:ACT_CON]->(pol)
with pol, length(path) + plen as plen, size(collect(distinct sc)) as sat_cons
match (pol)<-[:ACT_CON]-(rc)
with pol, plen, sat_cons, size(collect(rc)) as req_cons where req_cons = sat_cons

with plen, collect(pol) as pols order by plen asc limit 1
unwind pols as pol
return case when count(pol) = 0 or 'Deny' in collect(pol.decision) then 'Deny' else
'Permit' end as decision
```

# 4. A Sample Implementation Use-Case

We study a healthcare use-case example which involves patient, nurse, and doctor entities as well as digital assets such as patient profile and medical records. Peter is referred, by his family doctor, Sue, to a hospital for a few medical exams. John and Joe are working in the hospital as physician and nurse, respectively. The hospital provides digital services to maintain Peter's identity profile and his medical records, including a medical exam report, identified as "MR_1234".

## Implementing sample data in graph

The following graph presents the above data as a set of connected primitive and attribute nodes.

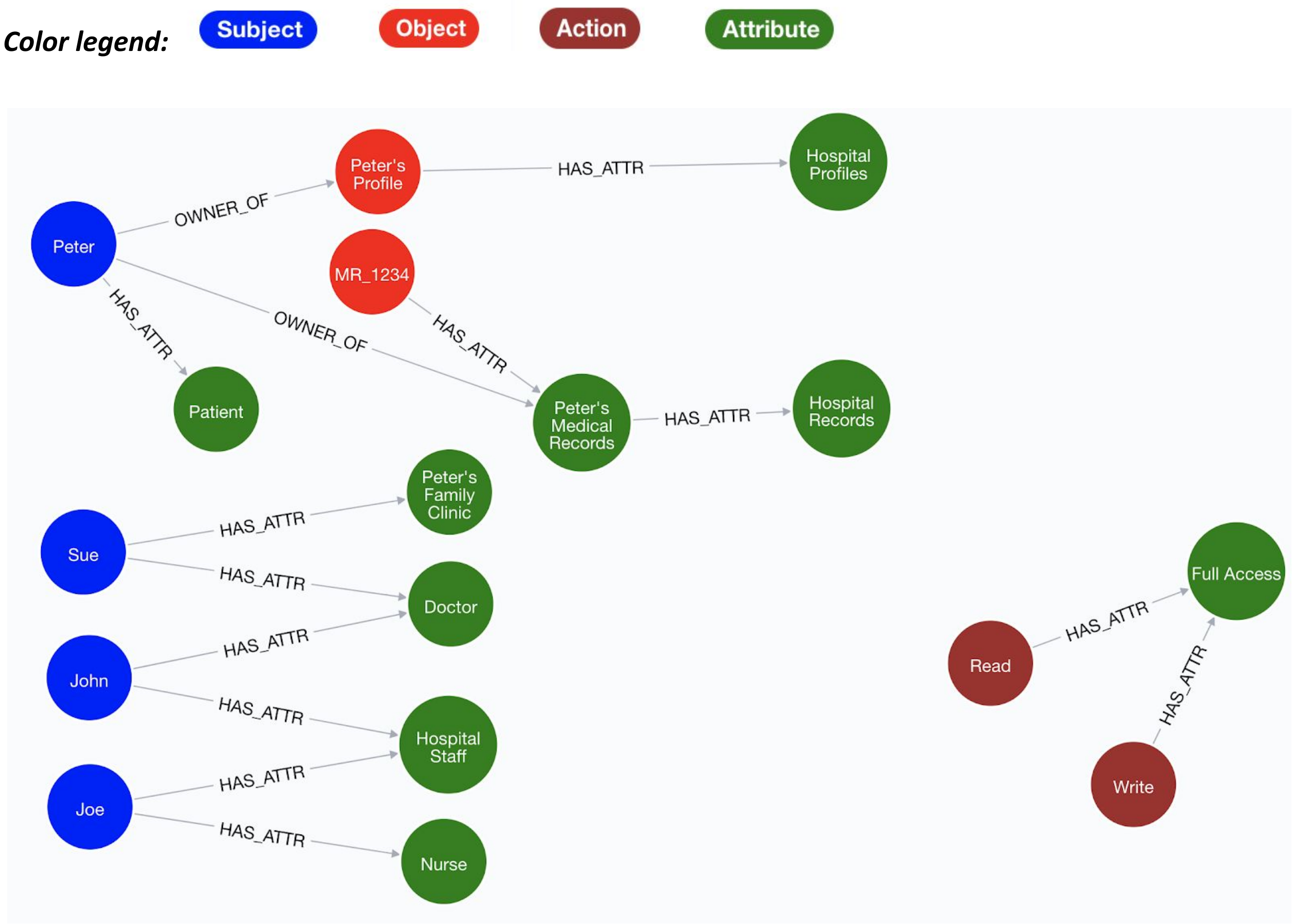

**Figure 1.** Neo4j graph presentation of the sample data

While there can be a variety of conceptual relationships between the primitive (subject, object, action) primitives and attributes, we use a separate "$HAS_ATTR$" relationship for those connections that are important to our ABAC policy matching. These distinguishable relationships help us define ABAC policies by connecting policy nodes to attributes nodes and processing runtime access queries by finding graph traversal pattern that connect query primitives to policy nodes via "$HAS_ATTR$" relationships. This relationship type can be viewed as part of the existing graph data or an ABAC-deployment graph layer. Appendix A shows the Cypher query that creates the above graph in Neo4j.

### 4.1. Sample access policies and graph implementation

The policies we are considering are:

- **Policy 1.** The hospital staff have full access to the hospital (patient) profiles.
- **Policy 2.** The hospital doctors have full access to the hospital records.
- **Policy 3.** Doctors from Peter’s family clinic can read Peter’s medical records.

As discussed earlier, these policies can be represented, respectively, as

- **Policy 1: *Permit if***
  - Subject is/has attribute “Hospital Staff”
  - Action is/has attribute ”Full Access”
  - Object is/has attribute “Hospital Profiles”
- **Policy 2. *Permit if***
  - Subject is/has attribute “Doctor”
  - Subject is/has attribute “Hospital Staff”
  - Action is/has attribute ”Full Access”
  - Object is/has attribute “Hospital Records”
- **Policy 3. *Permit if***
  - Subject is/has attribute “Doctor”
  - Subject is/has attribute “Peter’s Family Clinic”
  - Action is/has attribute ”Read”
  - Object is/has attribute “Peter’s Medical Records”

Policy 1 is a simple policy example with one condition per query primitive. Policy 2 extends this to a case where we account for multiple conditions of the same type (i.e., subject here). Policy 3 extends this further to a relationship-based policy which requires the subject to have a specific relation to the object; hence, attribute is a relationship. Figure 2 shows the data together with Policies 1, 2, and 3, shown by light blue nodes from top to bottom, respectively.

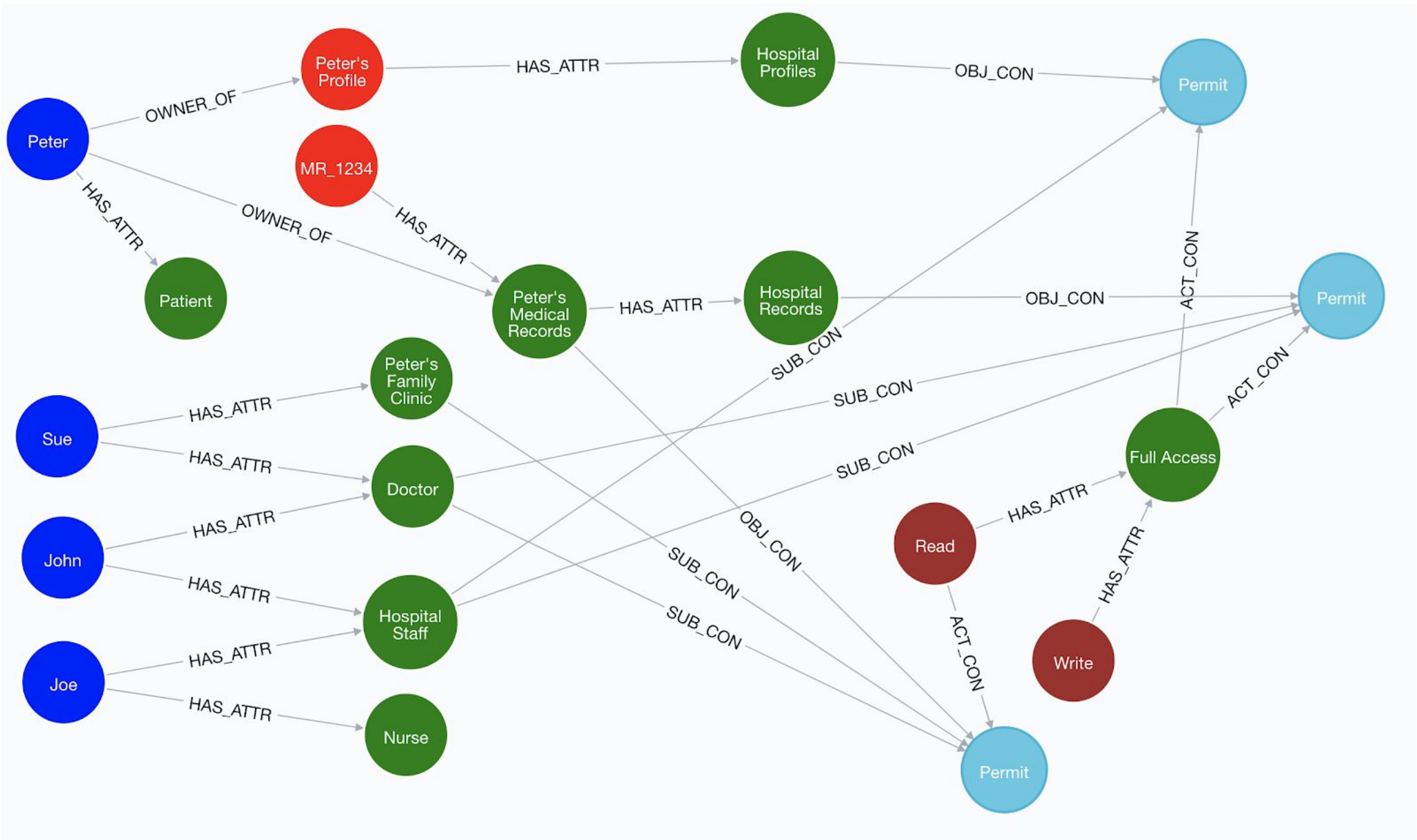

**Figure 2.** Neo4j graph presentation of the sample policies and data

Appendix B shows the Cypher query statements which create the three policy subgraphs in Neo4j.

## 4.2. Sample access queries and policy implications

Example implications of our sample policies include:

- John can write to Peter's medical record MD_1234.
- Sue can read MD_1234.
- Sue cannot write to MD_1234 record.

To examine the above, we write the following parameterized version of our "universal policy decision query" of Claim 1 with a "deny-overrides" combining algorithm:

```
with {AQ} as req
// Stage 1 - Subject Conditions
match (sub:Subject
{name:req.SUBJECT_NAME})-[:HAS_ATTR*0..5]->(sc)-[:SUB_CON]->(pol:Policy)
with req, pol, size(collect(distinct sc)) as sat_cons
match (pol)<-[:SUB_CON]-(rc)
with req, pol, sat_cons, size(collect(rc)) as req_cons where req_cons = sat_cons
// Stage 2 - Object Conditions
```

```
match (obj:Object {name:req.OBJECT_NAME})-[:HAS_ATTR*0..5]->(sc)-[:OBJ_CON]->(pol)
with req, pol, size(collect(distinct sc)) as sat_cons
match (pol)<-[:OBJ_CON]-(rc)
with req, pol, sat_cons, size(collect(rc)) as req_cons  where req_cons = sat_cons
// Stage 3 - Action Conditions
match (act:Action {name:req.ACTION_NAME})-[:HAS_ATTR*0..5]->(sc)-[:ACT_CON]->(pol)
with req, pol, size(collect(distinct sc)) as sat_cons
match (pol)<-[:ACT_CON]-(rc)
with req, pol, sat_cons, size(collect(rc)) as req_cons where req_cons = sat_cons
return case when count(pol) = 0 or 'Deny' in collect(pol.decision) then 'Deny' else
'Permit' end as decision
```

where `AQ` refers to an access query from the following parameters object:

```
:params { "JohnWriteMR_1234": { "SUBJECT_NAME": "John", "OBJECT_NAME": "MR_1234",
"ACTION_NAME": "Write" }, "SueReadMR_1234": { "SUBJECT_NAME": "Sue", "OBJECT_NAME":
"MR_1234", "ACTION_NAME": "Read" }, "SueWriteMR_1234": { "SUBJECT_NAME": "Sue",
"OBJECT_NAME": "MR_1234", "ACTION_NAME": "Write" } }
```

Running the Cypher with the three access queries, we will obtain:

- JohnWriteMR_1234 ⇒ **Permit**
- SueReadMR_1234 ⇒ **Permit**
- SueWriteMR_1234 ⇒ **Deny**

which matches our expectation.

# 5. Conclusion and Future Work

In this paper, we introduced a graph model to represent ABAC policies and provided a universal Cypher policy decision statement that decides about access queries by traversing the paths between query primitives and policy nodes. For the sake of simplicity, we assumed:

- Policies don't have negative or multi-level conditions.
- Attributes are defined only for “subjects”, “objects”, and “actions” primitives.

In this section, we provide some directions to expand our graph model to support compound policies as well as extended attributes. We leave a more detailed analysis of this to future work.

## 5.1. Compound conditions/policies

ABAC generally allows for “compound” policies, where each policy condition can be a nested combination of other conditions in the form of disjunction (OR), conjunction (AND), and negation (NOT) logics. Firstly, we note that even the simple policy graph model allows for one-level compound logic, i.e.,

- Logical disjunction is achieved as each policy subgraph is independently created and we match “any” policy against a given query primitive.
- Logical conjunction is implicitly supported as each policy node is of a conjunctive (AND) operation nature. We thus can connect multiple conditions of a type to a policy node.
- Logical negation is supported at the policy node level through the node property $decision \in \{Permit, Deny\}$.

This implies that our simple policy model is “complete” in the sense that any set of policies in any access management scenario can be written in the simple policy model. However, it is reasonable to assume that some access management scenarios desire managing more complex policies.

***Non-monotonic policies and negation nodes.***
Non-monotonic policies depend on conditions which require “lack” of an attribute for a subject/object/action primitive. An example policy is:

*“Any Employee whose account is NOT suspended can browse through company portal.”*

The above statement can be presented by a “$decision = Permit$” policy node with the following conditions:

- subject (is/has attribute) Employee, and
- subject (does NOT have attribute) Suspended, and
- action is (is/has attribute) Browse, and
- object is (is/has attribute) Company Portal.

Our simple graph model can support the above policy if is broken down (for example) to the following two simple policies and apply the “deny-overrides” combining algorithm:

- **Policy 1:** Permit if
  - subject (is/has attribute) Employee, and
  - action is (is/has attribute) Browse, and
  - object is (is/has attribute) Company Portal.
- **Policy 2:** Deny if
  - subject (is/has attribute) Suspended, and
  - action is (is/has attribute) Browse, and
  - object is (is/has attribute) Company Portal.

We can also expand our graph model to deal with negation in condition via explicit negation nodes that connect policy nodes to condition nodes, like:

$$(c) - [: con] -> (n : NOT) - [: con] -> (pol).$$

Here is an example of a subject condition graph:

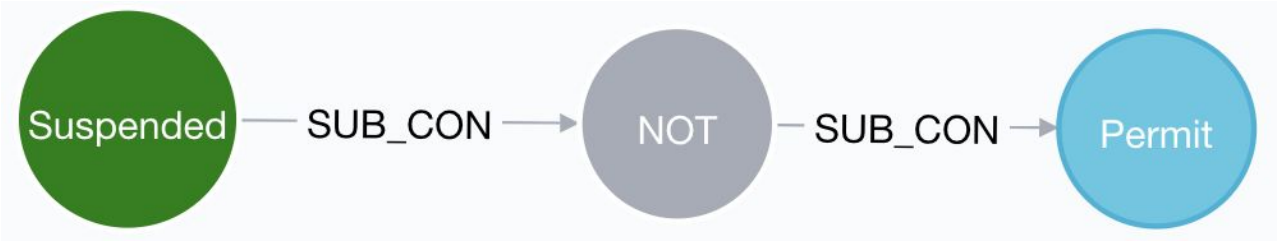

**Figure 3.** Negated subject condition

Lack of attribute for a primitive can be presented over graph as non-existence of a path between the attribute node and the primitive node. To support negation conditions within the universal decision query, the Cypher pattern between a negation condition node $c$ and a query primitive $x \in AQ$ should return "null" if $c$ is required by the policy node.

***Multi-level conditions and conjunction/disjunction nodes.***

An example of a multi-level condition policy is:

*"Any Manager or Senior Employee can View Monthly Reports."*

For our simple policy model to implement the above policy statement, we would need to break it down to two policy subgraphs as follows:

- **Policy 1:** Permit if
  - subject (is/has attribute) Manager, and
  - action is (is/has attribute) View, and
  - object is (is/has attribute) Monthly Reports.
- **Policy 2:** Permit if
  - subject (is/has attribute) Employee, and
  - subject (is/has attribute) Senior, and
  - action is (is/has attribute) View, and
  - object is (is/has attribute) Monthly Reports.

Although the above may look cumbersome for the above statement, staying with simple policy model and avoiding multi-level conditions may be a good practice from a performance point of view. If needed, however, to deal with multi-level policies in graph, we can use conjunction/disjunction nodes that combine other conditions. Figure 4 illustrates a multi-level condition policy subgraph for the original policy:

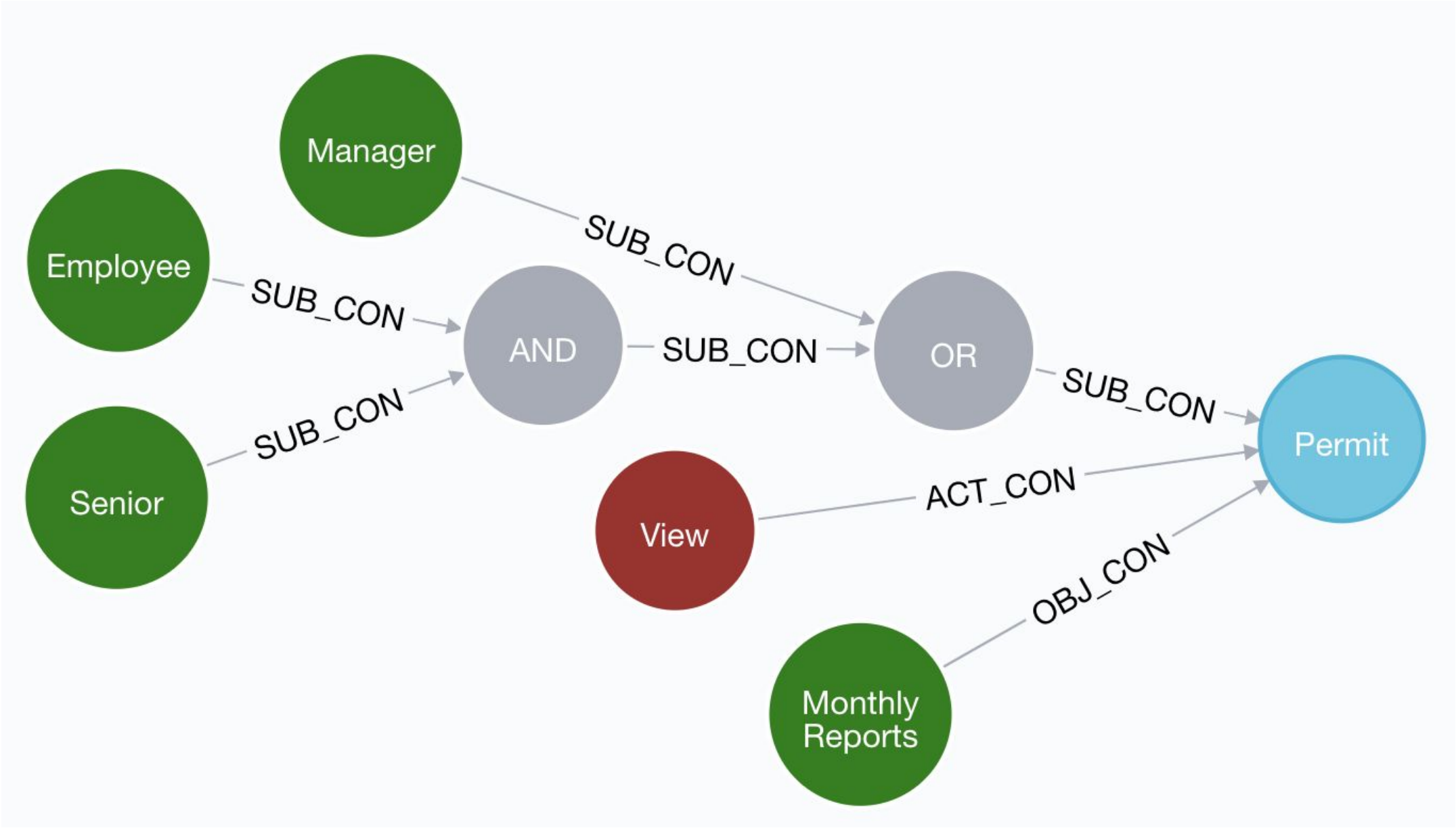

**Figure 4.** Policy with multi-level compound subject condition

## Extended attributes - session and environment

In this paper, we considered attributes only for “subjects”, “objects”, and “actions”. In generic ABAC, more primitives can be involved. We can extend “subject” to “session”, i.e., which represents the requester of a runtime access and includes more than only subject:

- Subject
- Location
- Client Type
- Context (AuthN level/type, Creds, etc)

For example, a “session” can reveal that the request is from:

*“jsmith” with “iPhoneX App” in “Chicago” where “2-factor AuthN” is established*

Furthermore, ABAC can involve environmental attributes, which describe environment parameters such as current time, current temperature, etc. We note that our graph model can easily be extended to support attributes for session and environment primitives. While static information, e.g., location, time range, IP range, can be a good fit for the graph model, it is recommended to not to use graph for dynamic environmental primitives such as time, IP, etc.

# 7. Appendix

## Appendix A - Cypher to create sample data

```
//Patient
create (peter:Subject:User:Primitive
{name:'Peter'})-[:HAS_ATTR]->(patient:Role:Attribute {name:'Patient'})

//Doctor & Nurses
create (joe:Subject:User:Primitive {name:"Joe"})-[:HAS_ATTR]->(nurse:Role:Attribute
{name:'Nurse'})
create (john:Subject:User:Primitive
{name:'John'})-[:HAS_ATTR]->(doctor:Role:Attribute {name:'Doctor'})
merge (john)-[:HAS_ATTR]->(staff:Group:Attribute {name:'Hospital
Staff'})<-[:HAS_ATTR]-(joe)
merge (sue:Subject:User:Primitive {name:"Sue"})-[:HAS_ATTR]->(doctor)
merge (sue)-[:HAS_ATTR]->(pClinic:Relationship:Attribute {name: "Peter's Family
Clinic"})

// Peter's digital assets
merge (peter)-[:OWNER_OF]->(pRecs:Group:Attribute {name:"Peter's Medical
Records"})<-[:HAS_ATTR]-(rec:Record:Object:Primitive {name:"MR_1234"})
merge (pRecs)-[:HAS_ATTR]->(hRecs:Group:Attribute {name:'Hospital Records'})
merge (peter)-[:OWNER_OF]->(prof:Data:Object {name:"Peter's
Profile"})-[:HAS_ATTR]->(pp:Group:Attribute {name:'Hospital Profiles'})

// Actions
merge (read:Action {name:'Read'})-[:HAS_ATTR]->(fullAccess:Attribute:Group
{name:"Full Access"})<-[:HAS_ATTR]-(write:Action {name:'Write'});
```

## Appendix B - Cypher queries to create policies

```
// Policy 1 - Permit: Hospital Staff, Full Access, Patient Profile
```

```
match (sub:Attribute {name:"Hospital Staff"}), (obj:Attribute {name:"Hospital
Profiles"}), (act:Attribute {name:"Full Access"})
create (pol:Policy {name:'Policy1', decision:"Permit"})
merge (pol)<-[:SUB_CON]-(sub)
merge (pol)<-[:OBJ_CON]-(obj)
merge (pol)<-[:ACT_CON]-(act);

// Policy 2 - Permit: Doctor & Hospital Staff, Full Access, Hospital Records
match (sub1:Attribute {name:"Hospital Staff"}), (sub2:Attribute {name:"Doctor"}),
(obj:Attribute {name:"Hospital Records"}), (act:Attribute {name:"Full Access"})
create (pol:Policy {name:'Policy2', decision:"Permit"})
merge (pol)<-[:SUB_CON]-(sub1)
merge (pol)<-[:SUB_CON]-(sub2)
merge (pol)<-[:OBJ_CON]-(obj)
merge (pol)<-[:ACT_CON]-(act);

// Policy 3 - Permit: Doctor & Peter's Family Clinic, Read, Peter's Medical Records
match (sub1:Attribute {name:"Peter's Family Clinic"}), (sub2:Attribute
{name:"Doctor"}), (obj:Attribute {name:"Peter's Medical Records"}), (act:Action
{name:"Read"})
create (pol:Policy {name:'Policy3', decision:"Permit"})
merge (pol)<-[:SUB_CON]-(sub1)
merge (pol)<-[:SUB_CON]-(sub2)
merge (pol)<-[:OBJ_CON]-(obj)
merge (pol)<-[:ACT_CON]-(act);
```

## Appendix C. Proof of Claim 1

*We need to show:*

A. *For each returned policy, all its required conditions are satisfied by* $AQ$.
B. *Any policy whose required conditions are all satisfied by* $AQ$ *is returned.*

*Let us prove (A) first. Assume* $pol$ *is a returned policy by the above statement. We show that all required conditions of* $pol$ *of type* $SUB_CON$ *are satisfied (similar argument holds for other condition types). Assume that there is a required condition* $c$ *that is NOT satisfied by query primitive* $sub$. *This would mean (via Definition 3)*

$$(sub)-[:HAS_ATTR*0..5]->(c)$$

*is empty. Comparing the first two lines of the Cypher statement, we would have:*

- *Size of* $sc$ *is at most equal to the size of* $rc$ $(size(sat_con) \leq size(req_con))$
- *Node* $c$ *is part of* $rc$ *but not part of* $sc$.

*This would imply $size(sat_con) < size(req_con)$ which violates the Cypher condition clause, leaving "$pol$" out of the returned policies; hence, proven by "**contradiction**".*

*We now prove (B). Assume $pol$ is a matching policy $pol$ for $AQ$. All required conditions $rc$ of $pol$ of type $SUB_CON$ are satisfied by $sub$, implying*

$$(sub) - [: HAS_ATTR * 0..5] -> (c) - [: SUB_CON] -> (pol)$$

*holds for all $c$ in $rc$, meaning $rc$ is a subset of $sc$. On the other than, the first two lines of the statement suggests that $sc$ is a subset of $rc$. This together proves the equality $size(sat_con) = size(req_con)$; hence $pol$ survives the subject condition check. Similarly, we can show that $pol$ survives other condition checks and hence is returned by the Cypher statement.*